# Alive publication

Mikhail Gorbunov-Posadov

Keldysh Institute of Applied Mathematics, Russian Academy of Sciences

gorbunov@keldysh.ru

**Abstract.** An alive publication is a new genre for presenting the results of scientific research, which means that scientific work is published online and then constantly developing and improving by its author. Serious errors and typos are no longer fatal, nor do they haunt the author for the rest of his or her life. The reader of an alive publication knows that the author is constantly monitoring changes occurring in this branch of science. Alive publication faces the inertia of scientific publishing traditions and, in particular, traditional bibliometrics. Unfortunately, at present, the author who supports an alive publication is dramatically losing out in many generally accepted bibliometric indicators. The alive publication encourages the development of the bibliography apparatus. Each bibliographic reference will soon have to contain such important for the reader updating on-the-fly attributes as attendance, number of external links, date of the last revision, etc. It is to be expected that as the alive publication spreads over to the scientific world, the author's concern for the publication's evolution will become like a parent's care for the development of a child. The Internet will be filled with scientific publications that do not lose their relevance over time.

## Contents

# 1. Introduction

An alive publication is a scientific work published on the Internet that is constantly being developed and improved by its author. This genre of scientific publication has a number of obvious advantages.

- Author who has abandoned traditional, static publication in favor of alive publication find themselves in a new, much more comfortable, and productive environment. Serious errors and typos are no longer fatal, nor do they haunt the author for the rest of his or her life. The circle of readers of an alive publication is much wider. Interest in the publication often even increases over time; many readers return to their favorite text, not only to refresh their memory of the most significant moments but also to learn how the author's views are transformed and what new trends the author has noticed in the area under consideration.

- For the reader, an alive publication is undoubtedly preferable to a static one. Readers feel much more confident when they know that the text in front of their eyes is under the constant control of the author, who has carefully corrected all the errors and inaccuracies noticed since the first publication of the work online and is constantly monitoring changes occurring in this branch of science.

The following is a typical example of the advantages of alive publishing. In 2016, the last volumes of the Great Russian Encyclopedia (GRE) [ 1 ] were published, and the GRE was posted for open access online. In 2018, GRE site traffic was 90 thousand hits per month. The traffic to the Russian-language Wikipedia [ 2 ] in the same period was 900 million hits per month. Thus, for one appeal to a static GRE that quickly goes out of date, there were 10 thousand requests to the alive Wikipedia. Of course, "static/alive" is not the only difference between GRE and Wikipedia. Yet such dissimilar traffic figures seem to be primarily due to this difference.

However, an alive publication only partially resembles a Wikipedia article. The main difference is that in Wikipedia, all members of the wiki community edit the article and maintain the relevance of its content, but no one is personally responsible for the article. Therefore Wikipedia is often forced to put up with the existence of articles forgotten by the authors, the relevance of which has long passed.

In contrast, the author's name of an alive publication is explicitly placed next to the title. The author is the sole owner of the alive publication: he, and only he, has the right to make changes to it at any time. An author who has declared that his or her publication is alive can himself or, if the publication is commissioned, on someone's behalf assume the responsibility not only to constantly improve it but also to monitor developments in the research area and systematically reflect everything new in the text under consideration.

The use of the term "alive" publication is not conventional practice. The terms "living" [ 3, 4 ], "evolving" [ 5 ], "dynamic" [ 6 ], "liquid" [ 7 ], "propelled", "movable", "progressing", "developing", and "advancing" are used more often.

To our regret, most of these terms generally mean that the publication contains multimedia and/or interactivity rather than alive content.

## 2. Status

The attitude towards alive publications is as yet ambiguous. Recent fundamental reviews [ 8, 9 ] of prospective models of a journal article do not even mention alive publication. Nevertheless, the project "Wikiversity: Journal of the future" placed alive publication in the first principal position among the main features of the journal of the future:

> *Research is a process. The scientific journal of the future provides a platform for continuous and rapid publishing of workflows and other information pertaining to a research project, and for updating any such content by its original authors or collaboratively by relevant communities. Eventually, all scientific records should have a public version history or a public justification for not having one* [ 10 ].

The English version of Wikipedia does not have an article entirely dedicated to alive publication. However, in the Russian version of Wikipedia, such an article is present [ 11 ].

The life of an alive publication begins when the author posts it for the first time in an institutional archive, on a preprint server, or in another online repository. The main point is that this repository does not limit the number of subsequent revisions of the stored publication.

Along with the development of the content, other next steps are also possible. In particular, the author then thinks about the status of this alive publication.

The status of a scientific article is largely determined by obtaining the approval of reviewers of a reputable journal. Therefore, the author of an alive publication usually sends one of versions of the text to some journal. An alive publication feels most comfortable in an overlay journal [ 12 ], where the full text of the article is posted in a repository independent of the journal. The repository usually supports content elaboration. Subsequently, the alive text posted online is supplemented with a link to the journal. This link serves as a guide for the online reader in terms of the article quality.

The reader of subsequent versions, of course, is aware that the journal reviewed not the most recent but some earlier version of the text, which may have been preserved somewhere in the protocols but is no longer of interest. However, the mention of a journal publication usually looks like a reliable sign of quality in the eyes of the reader. It seems unlikely that the author wrote a good-quality article, went through the thorns of peer-reviewing in the journal, but then worsened the quality of the text as a result of subsequent editing.

The connection between an alive publication and an overlay journal is built very organically. The overlay journal does not print or copy the full text of an alive publication to its site but only places a link to the publication's position in the source repository. It is desirable for the overlay journal, in turn, to be alive. In such a case, the information, abstract, review, or illustration included in the overlay journal evolve along with the evolution of the alive publication. The union with the alive overlay journal is not broken even when an alive article gradually transforms into an alive monograph as there are an increasing depth and volume of the alive content.

Certain difficulties arise in the administrative bibliometric evaluation [ 13 ] of an alive publication. In particular, it is not clear what should be specified as the year of publication in this case. Alive publications are often supported over many years, taking up a significant part of the author's working time. Therefore, when answering the question about the number of publications in recent years in a questionnaire, the author may be a notable loser.

The authors of alive publications often lose out to their colleagues in traditional bibliometric indicators because instead of writing several publications, they are engaged in improving only one. There may be a lot of links to an alive publication; however, there are not many alive publications themselves, which is why, for example, their Hirsch index (h-index) can lose significantly.

However, some bibliometric indicators present an alive publication in a favorable light. For example, a well-known altmetric is the number of bookmarks saved by site visitors. Visitors to a traditional publication save bookmarks for well-known reasons: to refer to the publication in their works, or to reread it if something is forgotten. In an alive publication, an impressive addition to these reasons is the reader's interest in the author's new results and in the recent developments in the field under consideration reflected in the alive publication.

So far, unfortunately, using the number of bookmarks as a full-fledged altmetric is not recommended. A reference management system can count bookmarks. However, there are too many similar systems, such as Mendeley [ 14 ], EndNote [ 15 ], Zotero [ 16 ], and others. Wikipedia [ 17 ] counts approximately 30 such systems. A single bookmark counter for all systems has not yet been implemented, so this interesting altmetric is used infrequently.

## 3. New genre

Let us consider a few typical features of the genre of alive publication.

An author who has published a traditional static article and then received a new result in the same field is forced to prepare another additional article. Like any other article, the additional article must be self-contained. Therefore, a significant part of it must be allocated to repeating the conclusions obtained in the original article. This leads to a need to duplicate texts, which interferes with the reader who is already familiar with the initial article. In addition, because of this inevitable duplication, the author may be reproached for self-plagiarism [ 18 ]. The author of the alive publication is spared from these troubles. This author does not duplicate anything but simply adds new results to the existing alive text.

At different times an alive publication may receive reviews from competent experts. The number of reviews is unlimited. The author, of course, reflects the reviewer's considerations in the alive text every time. Thus, the reviewer becomes, to some extent, a co-author of the publication.

An alive publication can be accompanied by a blog on which everyone can express an opinion about it. The author actively participates in this discussion and constantly reflects the discussion results in an alive text. Here, the evolution of the alive publication begins to resemble crowdsourcing.

Technically, working in the alive publication genre is easy. Many scientists have already mastered these skills by keeping their personal web page or profile up-to-date. Here, the scientist must constantly update their scientific biography and main results. If a personal web page includes the professional credo, then the continuous reflection of its evolution is essentially an alive publication.

It would be useful to convert many types of scientific work into the genre of alive publication. For example, it makes sense to make corrections to the online text of a thesis after the defense to reflect the views expressed during the discussion. Then, for several years, the new Philosophiæ Doctor could keep this text up to date, reflecting both the author's new results and changes occurring in the research area. Of course, the original text of the thesis discussed during the defense must also be stored and available.

Each major research center in a specific area could initiate the creation and regular maintenance of an alive specialized encyclopedia. This task is both vital and honorable. Compiling a vocabulary and allocating responsibilities for maintaining articles between individual specialists is performed in cooperation with related scientific organizations. The main point is that all authors of an individual article consider it their most important duty to constantly maintain the up-to-date text, reflecting the evolution of this scientific field.

The appearance of alive publications has significantly transformed many of the usual connections in the infrastructure of scientific publications. For example, in a traditional publication, all elements of the bibliographic list were published before it was published, and all articles that quote it are published after it was published. In contrast, an alive publication often refers to an article published after the first appearance of this alive publication. In addition, an article quoting an alive publication may come out long before the current version of the alive publication appears. Thus, the existing bibliometric indicators require a certain correction.

## 4. Immortality

What happens to an alive publication when the author, for some reason, loses the ability to systematically keep it up to date?

An alive publication can be, for example, an article in an alive specialized encyclopedia. The care of such an alive publication was probably entrusted to the author by some research center that maintains this alive encyclopedia on its website. If the author is now out of the game, the research center must find a successor to the author who will inherit the responsibilities for this publication.

If you, as the author of an alive publication was maintaining it on your own initiative, then you certainly has the right to bequeath its subsequent updating to someone. In this case, the publication does not just continue to exist but also retains its "alive" status. If you does not have a successor, the future fate of this alive publication seems to be no different from that of any other publication published on the Internet.

Of course, it is necessary to think about safety. In 2013, the G8 Science Ministers published a corresponding statement:

> *We recognize that the long-term preservation of the increasingly digitized body of scientific publications and data requires careful consideration at the national and international levels to ensure that the scientific results of our time will be available also to future generations* [ 19 ].

As long as an alive publication is truly alive, backing it up requires special attention. The point is that when a modification is made to an alive publication, we should probably be made simultaneously the same change to the backup copy. If the publication becomes static, it is copied as usual.

In the WebCite [ 20 ] and Perma.cc [ 21 ] projects, an interesting approach is suggested in which the care for the backup of an online publication is assigned not to its owner but to the author of the work that links to this publication. Before posting a hyperlink to someone's article, publishers who follow this approach copy this article to the reserving depository. The hyperlink, in this case, contains two addresses: the direct address of the publication and the reserved address of its copy in the depository. Unfortunately, these projects do not serve alive publications since such copies of alive articles do not track their changes.

Also, of course, it is necessary to ensure that when the publication address is changed (which often happens when the site is reorganized), external links are not affected. The DOI (digital object identifier) will help here. The publication can be assigned a unique number that replaces the explicit URL when linking to it. The URL that indicates where the publication is currently located is mapped to the number in the DOI project tables. If the publication address changes, then the URL is changed in the DOI table, and all links to the publication are correct again.

For a long time, an alive publication could not access DOI services. The main DOI registrar for scientific publications is Crossref. Alive publication and Crossref were incompatible since Crossref required strict immutability of the object that received the DOI. However, the number of alive publications steadily increased, and a few years ago, the requirement for total invariability of a publication that received a DOI was removed.

Unfortunately, most of the leading scientific journals did not notice the removal of the total invariability requirement. To this day, they do not allow changes to be made to the online version of an article, even to correct an error that was noticed [ 22 ]. This is simply monstrous. Fixing an error in the online version requires a few minutes of work; all subsequent online readers would be able to read the correct text. It is absurd to extend ideas about the immutability of a printed article to the Internet.

## 5. Support for changes

As noted above, publishers are dominated by a centuries-old tradition of paper publications, where the published text was perceived as a once-and-for-all formed monolith that resolutely does not allow any alterations. In particular, the existing mechanisms of bibliographic references, citations, and many other connections commonly used in the world of science are guided by the idea that any publication is written in stone. This apparently largely explains the unnatural order for the Internet, according to which the vast majority of scientific journals that post published articles online categorically do not allow their authors to change anything later.

Indeed, if we allow published texts to change, we must transform some of the established views. For example, if special means of servicing alive publications is not provided, a fragment of text quoted by someone from outside may change or disappear altogether, or previously expressed criticisms of the work may hang in the air because the author of the alive publication has improved or corrected its text.

In [ 23 ], a fundamental approach is proposed to ensuring that the reference to an alive publication is correct. If an article referring to an alive publication is itself assumed to be an alive article, then it is possible to avoid the disagreement of this reference when making changes using a specific tool. The link to an alive publication is encoded in a special way; thus, when the cited material changes, all authors of works that reference to it receive notifications, where their references are marked as possibly outdated. In response to this notification, the reference author analyses the changes that have occurred and perhaps correct the link; this reference becomes thereby relevant again.

Special consideration should be given to peer-reviewing an alive publication. Is it possible to mobilize a reviewer who will promptly monitor every change made by the author of the publication and allow or not allow its appearance on the official website of the publication? This would probably be too difficult and time-consuming.

The problem is easily solved if the publisher implements the simultaneous existence of two versions of the publication on the site: the official version that has passed review and the author's version, which has not yet been approved by the reviewer. Then, the decision is up to the site visitors: they choose which version they would like to see.

The status of the author's version here becomes similar to the status of a site visitor's comment on online work. The editorial board can pre-moderate both comments and changes and can block extremist statements, but the board is not responsible for their quality or content. Of course, at the request of the author, the editorial board can conduct one more peer review in transferring the author's version to the official status. However, the author should not abuse such an opportunity. For example, in the Ridero publishing house [ 24 ], changes introduced to an alive print-on-demand book are checked and recorded no more than once a quarter. The arXiv provides the author more freedom: "We ask that articles be replaced no more than once per week" [ 25 ].

## 6. Versions

A popular and reliable support scheme assumes that changes made by the author to an alive publication are always logged in the form of versions (generations, revisions). In this case, in particular, criticisms concerning the previous state of material corrected later do not lose their meaning if they are addressed to an earlier version of the text that was saved in due time. The protocol of changes is useful also for providing related comments from the author, allowing regular readers to easily track all the innovations that appear in the text.

The forms of version support of an alive publication are not yet well established. Let us consider typical solutions using the example of well-known systems.

Crossref is a DOI registration agency that has been running the version support service Crossmark [ 26 ] since 2012. According to Crossmark, the "Check for updates" button (Figure 1) is eye-catching and prominently displayed in an online document.

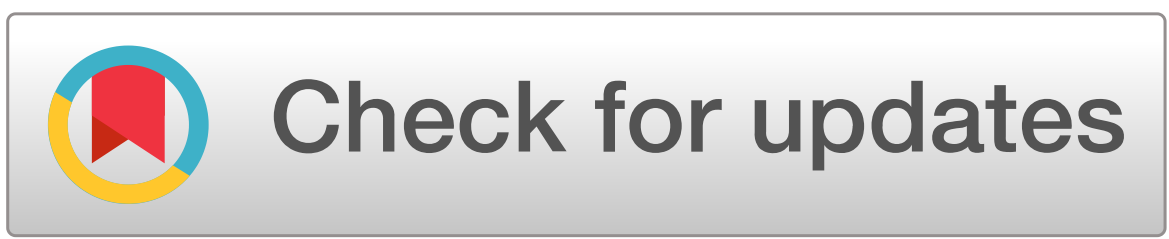

*Figure 1.* CrossMark button supporting versions of a publication in Crossref

By clicking on this button, the reader can determine whether a more recent version of the material exists and where it is located as well as whether this material has been retracted by the editor.

F1000Research [ 27 ] is an open-access publishing platform. In F1000Research, when you encounters an early publication text that already has an updated version, the first thing you sees is a message in a pop-up modal window that this version is outdated and that there is an opportunity to read the more recent text. Of course, the vast majority of visitors who see such a message will go to the updated text, where at least the errors noticed will have been corrected. The F1000Research scheme seems more practical than the Crossref scheme, where a web page visitor may simply not notice the "Check for updates" button and may therefore read the outdated text.

However, both the Crossref scheme and the F1000Research scheme have drawbacks. When forming a bibliographic reference, the obsoleting version of the text should not be abused. An out-of-date version is generally not interesting to ordinary visitors. It may be important only to a historian of science or a lawyer who proves the precedence of an author. For the mass reader, it is much more comfortable to immediately access the latest version of an alive publication. It is only a rare fan of antiquities who may be interested in the history. For this enthusiast you can provide a special button, "Protocol of changes".

The electronic arXiv [ 28 ] overlay journal Sigma [ 29 ] supports the alive publications and refers primarily to the latest version of an alive article. If the author wants to make changes to an already published article, the journal publishes a new version of the article on its website and replaces the article in the arXiv. In the case of minor changes (when referees are not needed), the journal makes them quite quickly, within a few days. The previous version also continues to be stored, and it is assigned the same name with the addition of the suffix v<i>, where <i> is the version number. This means that the suffix can be used to refer to any version as well as to the latest version without the suffix. However, in Sigma, you cannot refer to the latest version with the suffix, which makes naming somewhat difficult.

The last drawback is missing from the version naming scheme used in arXiv. Here, the recent version has both a short name without the suffix and a name with the recent version number suffix. This scheme allows you to refer both to the alive publication as a whole and to any specific version of it, including the recent one. In 2022, publications in arXiv received DOI. The DOI is the same for all versions of an alive publication and always refers to the most recent one.

However, due to the imperfections of the current publication infrastructure, certain conflicts arise here as well.

For example, the third version of an article placed in arXiv [ 30 ] was published in 2018 in the traditional non-overlay journal "Classical and Quantum Gravity" [ 31 ]. A year later, the authors had placed a new fourth version of the article in arXiv (Figure 2). When publishing the third version of the article, the journal assigned the article a DOI and placed the Crossmark icon (Figure 1) on its web page. However, the traditional journal does not support alive publications, pretending that it knows nothing about the existence of the arXiv. Therefore, the journal did not reflect the appearance of the new fourth arXiv version in Crossref. As a result, the Crossmark icon confuses online readers of the third version of the article published in the journal by mistakenly informing them no more recent version existed.

| Journal reference: | Class. Quantum Grav. 35 065010 (2018) |
| DOI: | 10.1088/1361-6382/aaaafa |
| Cite as: | arXiv:1710.02185 **[gr-qc]** |
| | (or arXiv:1710.02185v4 **[gr-qc]** for this version) |

**Submission history**

From: LVC Publications [view email] [via Lvc Publications as proxy]
[v1] Thu, 5 Oct 2017 19:18:51 UTC (4,609 KB)
[v2] Wed, 11 Oct 2017 16:26:26 UTC (4,608 KB)
[v3] Tue, 7 Nov 2017 22:45:22 UTC (4,608 KB)
**[v4]** Tue, 8 Oct 2019 13:29:33 UTC (4,507 KB)

*Figure 2.* Four versions of a publication on arXiv

This error is a characteristic consequence of the imperfections of scientific publications' existing infrastructure, which still do not always support alive publications.

## 7. Preprint: morphological misunderstanding

Nevertheless, version support services for alive publications have become widespread. For example, in 2019, such support was implemented even on the pirate site Sci-Hub [ 32 ]. Gradually, the understanding has developed that the life path of a modern scientific article is a long chain of links "several generations, optional journal publication, several generations, etc.," with a journal publication being neither the main nor a mandatory link in this chain.

The alive publication is confidently moving towards leading positions. The fetishization of the publication of an article by a traditional journal is becoming a thing of the past [ 33, 34 ]. Back in 2002-2003, Gregory Perelman published his famous proof of the Poincaré conjecture not in a journal but in the form of an alive publication in arXiv [ 35, 36 ]. This did not prevent him from then proving his copyright and gaining worldwide recognition.

Currently, the actual printing of a publication may not happen, and publication may well be limited to an online form. It is usually more convenient for readers to become acquainted with an article that interests them not in a journal but in an institutional repository or a preprint archive. Only in such environments can they confidently expect to obtain access to the latest version of the alive article. There is a growing understanding that the most organic representation of a scientific publication is a long series of generations. It is time to stop dividing these generations into the pejorative "before the journal" (preprint) and "after the journal".

Publication in a journal is just an intermediate snapshot, an ordinary episode in the article's rich biography. At the same time, the event of publication in a popular authoritative journal is the most massive source of comments, suggestions, and development. However, it is here, at the most fruitful moment, that the vast majority of publishers end the life of an article. This behavior of publishers is an obvious crime against science.

To date, the most popular form of alive publication is a preprint. Morphology and semantics of the modern interpretation of the word "preprint" confuse everything. Part "pre-" here has long meant not only "pre-" (the preceding), but also the intermediate and subsequent. Part "print" in most cases is decisively replaced by "online".

Preprints are becoming more widely distributed. In 2016, Crossref opened a dedicated preprint service [ 37 ]. In particular, Crossref membership was allowed for preprint servers, and custom metadata was implemented that reflects the author's workflow from preprint to official publication and further.

Over the past few years, leading scientific journal publishers have opened preprint servers for their authors: Springer [ 38 ], Elsevier [ 39 ], IEEE [ 40 ], MDPI [ 41 ], etc. [ 42 ]. Some of these servers accept preprints only before publication in the journal. However, most servers continue to accept new versions of the article after the journal publication in the same way, ensuring the parallel coexistence of the worlds of traditional journals and alive publications

Unfortunately, the coexistence of a traditional journal and an alive publication cannot yet be called harmonious. While the servers of alive publications consider it their duty to place a link to an appearing journal publication, journals, by contrast, do not generally place links to subsequent versions of alive publications (Figure 2). The event of publication in a popular authoritative journal is a powerful source of comments, suggestions, and development. However, it is here, on the rise, that the vast majority of publishers cut the life of the article. The reader of the journal does not see the link to the latest version of the alive publication, issued in response to the new data. This kind of publisher behavior is clearly a crime against science.

The idea of the leading role of the journal is long gone. Two parallel spaces - alive preprints (more precisely, alive publications) and peer-reviewed journal articles - now exist on equal rights. It remains only to manifest mutual respect. Not only the preprint is obliged to mention the appearance of a journal article on its basis, but the article is also obliged to inform the reader about the existence of the preprint, which was published later and strengthened the article.

The radical solution - to allow changes to the online text of the article on the site of the scientific journal - is practically not found yet. The only hopeful wording that came across on a journal's website was "Changes can be made to a paper published online only at the discretion of the Editorial Office".

A decisive change in the perception of the status of preprints occurred in 2020 in connection with the COVID-19 pandemic. Here, the promptness of preprint release brought this approach into a leading position: among the countless number of scientific publications devoted to COVID-19, almost half came out in preprint form. MIT has created an overlay journal dedicated to Covid-19, which independently selects among the preprints and reviews the most important works on this topic.

## 8. Living bibliographic reference

With the transition of a scientific publication to online, it becomes possible to include in it some dynamic elements along with the traditional static printed content. Dynamism appears for two reasons. First, the publication may change, as we saw above, because the author decides to improve, correct, or update it. Second, certain elements included in the publication can be dynamic, changing automatically without the participation of the author under the influence of embedded algorithms.

To emphasize the difference between these two sources of dynamism, we will continue to call the publication modifying by the author "alive," but we will call an automatically modified element "living." In the future, we will need to apply a living element to serve an alive publication. Therefore, we now have to focus on selected living elements in more detail.

There is a growing understanding that modern scientific publication is primarily an online publication. If, as a result, the online version of a publication begins to be perceived as the main version and the printed version only as an auxiliary version, then the world of scientific publications will become much richer and more attractive. Authors and publishers will no longer consider it their primary duty to serve the reader of the printed version of the work. They will agree that the printed-version reader may be deprived of some of the acquisitions available to the reader of the online version. In particular, multimedia illustrations, direct access to databases, online calculations, etc., which are not available to the reader of the printed version, will become organic and natural elements of scientific publication.

There are only a few left modern printed publications that do not receive a full-fledged Internet projection. In contrast, there are more and more online scientific publications that no longer have any printed version. Forms of online representation of scientific publications are constantly developing and improving. At the same time, many design elements currently used for online publications are outdated and are explained solely by the inertia of print publications.

In particular, the design of an online bibliographic reference is forced to be a servant of two masters: WWW standards and print publishing standards. We must seek an interdisciplinary balance of their established traditions and interests. It is hard for print publishers to even start thinking in terms of pop-up hints, and the dynamic formation of a living text of a bibliographic reference requires excessive restructuring of their consciousness.

In addition, noticeable inertia has accumulated from the era of the formation of the Internet, when it was necessary to unrestrainedly save server calls, which led to poor online designs. The speed of information transfer on the Internet is constantly increasing, and it is time to make online bibliographic references richer, more informative, and more convenient for the reader.

The above-mentioned modern trend of transition from the PDF to the HTML format, which opens up interesting new opportunities for developing the apparatus of a scientific article, is fruitful. The trend is quite powerful, but we cannot yet talk about a general transition to HTML. At the same time, if we look closely at how bibliographic references are presented on the Internet, we will find that even if a publication as a whole is presented in PDF, its bibliographic list is often additionally duplicated in HTML format as well.

There are various reasons. PDF representation of a bibliographic reference is uncomfortable in terms of using copy & paste because extra line breaks appear in the copy. On the websites of many publishers, the full texts of publications are available only for a fee; however, the bibliography as the main source of bibliometrics is usually available for free, and in this case, it is presented as an auxiliary file in HTML format. Bibliographic references in the table of contents of an overlay journal [ 12 ] are usually presented in HTML format since a date of birth of the last revision of the current publication is highly desirable here. The HTML format allows each bibliographic reference to be clearly and effectively supplemented with a living number of articles that link to it (Figure 3).

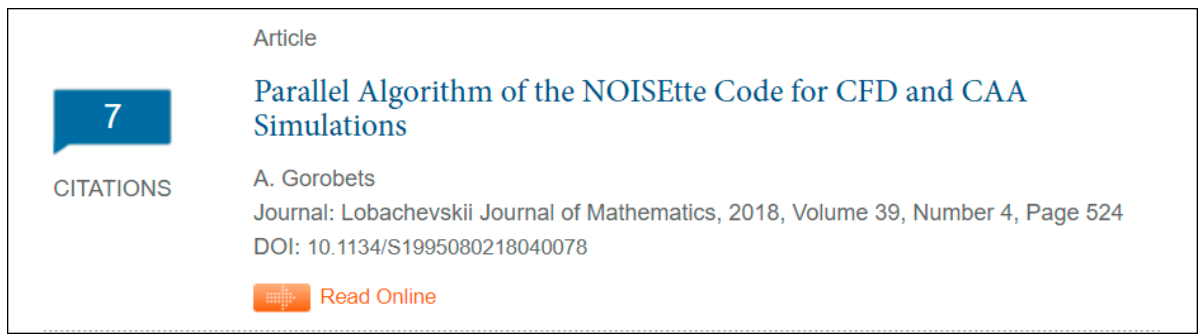

*Figure 3.* Bibliographic reference on the website of the Springer publishing house [ 43 ]. On the left is the living number of citations of the article

Anyway, there are already many HTML representations of bibliographic references on the Internet. There is no doubt that they are the future. Therefore, in many cases, it makes sense to discuss improving an online bibliographic reference in HTML without regard for its PDF representation.

One of the most attractive features of an online scientific publication is the use of hyperlinks, i.e., the ability to move from a bibliographic reference that interests the reader to the text of the scientific publication itself in just seconds. This feature is used extremely widely, but its potential has not yet been fully revealed. The composition of the information included in the bibliographic reference needs to be supplemented and improved utilizing data available through hyperlinks.

The potential of a hyperlink allows the composition of the attributes included in the bibliographic reference to be modified, orienting it to the online reader rather than the editor, publisher, librarian, bibliometrist, historian of science, etc. All these professionals can find the information they are interested in by clicking on the direct hyperlink to the online publication in question. At the same time, readers do not want to be forced to click on this hyperlink every time. They would like to get more useful information at an earlier stage. It is desirable that just by looking at the text of bibliographic reference, they will be able to make an informed decision—does it make sense to click on this hyperlink?

For example, readers are not interested in such traditional bibliographic components as the volume number and issue number of the journal or the range of page numbers of the article because they do not need to rummage through the library shelves and then turn the pages: a hyperlink will lead them directly to the beginning of the publication of interest. At the same time, an important reference point may be the number of this online article visits, i.e. an attribute that is not yet accepted as the significant component of a bibliographic reference.

The existing links in the modern Internet allow to dynamically, i.e., at the moment of the reader's request, build a list of "citing by," made up of bibliographic records of publications referencing this one. This opportunity is extremely useful because it constructively and clearly shows the direction of research developing the provisions set out in the cited publication. This is a living list that has no counterpart in the world of static print publications.

At one time, it seemed that a reverse bibliography list could be constructed by explicit Trackback [ 44 ] or Pingback requests being shipped to an online publication from newly appearing referencing documents. The online publication that received such requests collected them and thus formed its reverse list. Unfortunately, this interesting mechanism was soon ruined by spammers, and today, it is practically not used.

However, the usefulness of the reverse bibliographic list is not in doubt. Therefore, the lists appeared, but they are implemented otherwise. Reverse lists are built by many bibliographic systems based on the data they collect about publications. Now Google Scholar, Crossref, Web of Science, etc., allow you to obtain a living reverse bibliographic list.

The inclusion of a reverse bibliographic list in the information section of any online publication has already become the norm for leading foreign publishers. However, the bibliographic references in this list do not differ much from traditional ones. Of course, the usual view of a bibliographic reference is convenient for the reader. However, you quickly get used to the good. Therefore, if the bibliographic references in the reverse list become more informative, as suggested below, then the online reader will be grateful.

Both reverse and direct online living bibliographic references, especially if they are located in an HTML file, open up new opportunities. The entire text of bibliographic reference or some part of it can be generated based on the data available online and not written out explicitly. At first, you need to somehow get to appropriate online representation of the publication. Then you extract from this representation the meta-attributes of the publication and place them in the generated by you bibliographic reference. This decision eliminates errors that inevitably occur not only when directly rewriting text fragments but even when performing more reliable but still objectionable operations such as copy & paste. However, the most interesting thing here is that when the value of the publication attribute changes, this changed value also appears in the text of the bibliographic link to this publication.

The online representation can be either the source file itself, hosted in the primary repository, or other sources, such as Crossref meta-attributes, accessible via DOI of the publication. Increasingly, the information section of the source file includes links to files in the formats BibTeX, RIS, EndNote, Medlars, RefWorks, etc., which are quite capacious sets of publication meta-attributes.

Although interaction via the Crossref API looks more respectable, there are many arguments in favor of the primary file. First of all, a document mentioned in the bibliographic list may not have a DOI. In addition, Crossref does not support some document attributes that are interesting to the reader, such as the number of visits. Finally, any attributes of the primary file are entirely in the author's hands, while expanding the composition of meta-attributes in Crossref would require considerable effort.

Now, HTML, MS Word, and PDF have simple tools for declaring arbitrary attributes. It could be possible to standardize the representation of all meta-attributes that might be of interest to readers and place such attributes directly in the article file when creating or changing it. Whereupon, the need to explicitly specify these meta-attributes in the text of the bibliographic reference in some cases completely disappears. The URL or, even better, the DOI is specified and the appropriate API used to extract meta-attributes on-the-fly directly from the source file. Next, the extracted meta-attributes are combined in the HTML-text representation of the bibliographic reference in the desired manner.

As the experience of the well-known projects COinS [45], Dublin Core [46], etc. shows, direct placement of meta-attributes in the source file is quite feasible and entails no difficulties for either the publisher or the traffic. It allows you to eliminate duplication of mostly identical meta-attributes when indexing a publication in bibliographic systems. The author will have the opportunity to massively correct inaccuracies or errors in the header information because now corrections made in the source file will immediately be reflected in the text of all bibliographic references compiled in this way.

## 9. Living cross-domain communications

At the same time, it would be unwise to limit yourself to the materials of the source file of the publication when forming a bibliographic reference online. It makes sense to supplement the traditional set of meta-attributes with various living data that are interesting for the reader. Such data, which can be obtained through cross-domain queries, are listed below.

- The first thing that can significantly help the online reader is to check the validity of the hyperlink provided in the bibliographic link to the online publication. After all, it is no secret that the original text of the hyperlink may contain an error, or the long-term safety of the material posted online may be in question. Such a check can be successfully performed by the robot, eliminating the confusion of the online reader who encountered a broken hyperlink. Based on the robot's check results, a "broken" hyperlink can be excluded from the text or marked accordingly.

- If there is no hyperlink in the bibliographic link, this, generally speaking, does not mean that the material is not available online. The link could have been forgotten to include, or the material was posted with a delay, and at the time of preparation of this bibliographic reference, this material was still missing. To search for a hyperlink to an online posting of material, we can use, for example, the [47] service provided by Crossref. Adding the hyperlink found in this way to the bibliographic reference text will certainly please the online reader.

- The article that existed at the time the bibliographic list was formed could subsequently be retracted by the journal's editorial board. The generator of bibliographic reference text can learn about the retraction, for example, from the corresponding Crossref attribute. A message about retraction included on-the-fly in the bibliographic reference will markedly save the online reader time.

- The number of visits for the entire period since the publication was posted and for, say, the last 30 days is a useful [48] indicator that is not currently included in online text of bibliographic reference, likely because of the inertia of printed publications. For example, YouTube, which is not constrained by print restrictions, shows the number of views in a prominent place in its concise online link to the video (Figure 4). The source of information about visits (views) including in a bibliographic reference could be an analog of the project Counter [ 49 ], which would be turned not towards the publishers but towards the reader of bibliographic reference. Another possible source is the meta-attribute "Number of successful resolutions for the month," generated by Crossref but available now only in the monthly report [ 50 ] that Crossref sends to each publisher. However, Crossref recently complained about its generating statistics; it does not always cope with the elimination of robot visits [ 51 ].

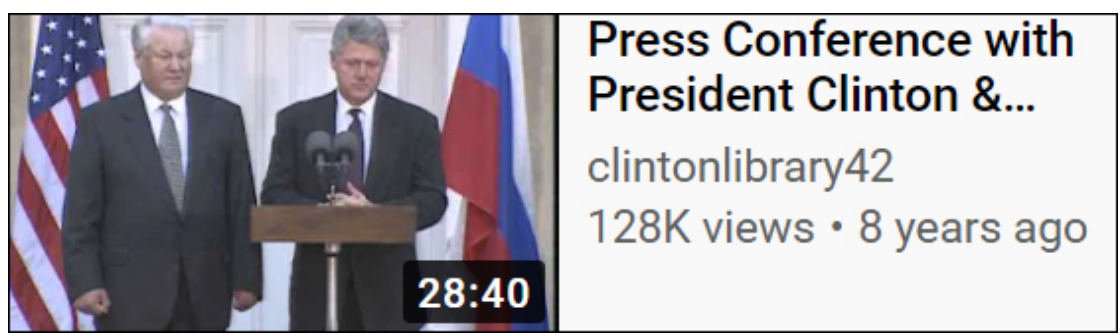

*Figure 4.* A reference to a YouTube video always includes
the living number of views

- The number of clicks (in total and/or in the last 30 days) on this bibliographic link from this online bibliographic list. Along with the specified above publication's traffic, this number of clicks can serve as a useful guide for the reader.

- Altmetrics—for example, the number of bookmarks for the publication made in reference management systems (Mendeley, EndNote, Zotero, CiteULike, etc.)—is a useful indicator.

- The living number of external links to the publication according to Google Scholar, Crossref (Figure 3), Web of Science, etc., is another useful indicator. Instead of such a living number of links, many publishers now supplement a bibliographic reference with static hyperlinks to records representing this publication in these bibliographic databases. By clicking on such a hyperlink, the reader will of course be able to find not only the number of external links but also much more information, in particular, what works have cited this publication according to information from the bibliographic database. However, it is still better not to force readers to click on hyperlinks unless it's critical; instead, the original bibliographic list should directly inform them about what is most interesting to them—about the living number of external links.

- Many publishers impose an open-access embargo at the beginning of a publication's existence, but after a certain period of time, the embargo is lifted. It is useful to inform readers dynamically about the current access mode directly in the bibliographic list. For a publication that has a DOI, this kind of verification is perfectly implemented in the Unpaywall project [ 52 ]. Further, if we find that access to the publication is not yet open, we can find out and offer the reader the address of the full text of this material on the popular pirate resource Sci-Hub. Of course, such a legally vulnerable hint could result in strong objections from commercial publishers.

- Translated version(s) of the article usually appears only sometime after the publication of the original version. In this case, it is possible (from the attributes of the source file or thanks to the Crossref relations "isTranslationOf" and "hasTranslation" [ 53 ]) to determine on-the-fly whether the translation(s) has appeared and inform the reader about it.

- Let the online publication can receive open reviews. Their list is publicly available and dynamically updated. Then the bibliographic reference can include information about the existence of a recent (say, no later than the last six months) review of the publication.

- Etc.

The inclusion of living information in the text of a bibliographic reference will make it significantly more informative and interesting for the reader. Depending on the load of the Internet channel and the server, dynamic attributes can either be created "on-the-fly," i.e., in the process of displaying the text of the bibliographic reference, or all the dynamic information stored on the server can be periodically updated, say, once a day (during a quiet time of night), to form actual texts of bibliographic references.

## 10. Living reference to alive publication

How can an alive publication be distinguished from a static one? Simply adding a special "Publication declared alive" icon to its web view is obviously not enough. After all, the author could place this icon once and then forget about it and his or her online text. Therefore, the only reliable evidence that the publication is alive is the fresh date of its recent revision. This date is placed prominently on the web page and serves as reliable guidance for the reader.

Along with it is desirable for an alive publication to be noticed earlier when viewing the bibliographic reference leading to it. How, for example, can a reader viewing a bibliographic list distinguish an alive publication included in it from neighboring static publications? This can be done by adding a living construction of the form "Last updated $\approx$ 2023-02-15 $\approx$" to the bibliographic reference, where the characters "$\approx$" frame the date of the last revision.

Here, support tools are essential [ 54, 55 ]. Because it is impossible to imagine an ultra-conscientious author who constantly looks through the bibliographic lists of his or her online articles to correct the changing update dates of alive publications every time.

When designing an alive publication, it is necessary to form the date of its last revision in a special way so that this date is available not only for the reader of the publication but also for other online articles that reference this publication.

The author of an online article forms a living reference to an alive publication in his or her text using support tools so that the date of the last revision in such a reference is automatically updated every time someone visits a web page. As already mentioned, to make it easy for the reader to notice the date of interest, this date is surrounded by the characters "≈", for example:

Gorbunov-Posadov M.M. Dynamically updated alive publication date // Publications, 2022, Vol. 10, No. 4, 48. Revision from ≈ 2023-01-04 ≈. https://preprints.org/manuscript/202209.0202

Citations within the text also require attention. If the publisher uses a numbering (Vancouver) system for citations (citations of the type "[1]"), then such citations are fully suitable for alive publication. If the Harvard method is used (style "author-date"), then to avoid misunderstandings, it makes sense to enlarge the citation to the following living format "(Gorbunov-Posadov, 2007, ≈ 2022-12-18 ≈)", i.e., (<author>, <the year of the first appearance of the work online>, ≈<generated on-the-fly date when the latest revision of the alive publication appeared online>≈).

For on-the-fly updated data in reference to an alive publication to become possible, two conditions are necessary. First, this date must be stored in the file of the alive publication and should be available for software. Second, the file with a reference to an alive publication must allow the implementation of inserting this date.

As mentioned above, tools for assigning values to arbitrary meta-attributes and, in particular, to the date of the last revision of the article are provided in all popular file formats of scientific publications: PDF, HTML, and MS Word. However, unfortunately, including this meta-attribute directly in the text of a living reference to an alive publication, is not always possible. Currently implemented support tools successfully update the date of the last revision only if the article containing the reference is presented in HTML format.

It should be reminded there are many HTML representations of bibliographic references on the Internet. There is no doubt that over time, there will be even more of them. Therefore, the tools for supporting a living date of alive publication can be of practical use right now, and their scope will only expand.

Unfortunately, examples of the use of an alive publication's living date in famous journals are not yet visible. The author of this article made several attempts to suggest to journals in the online HTML-version of his article that appropriate means be provided to support such dates. On each occasion, however, he received polite but firm rejections.

## 11. Conclusion

What can be expected in the area of alive publications?

The coronavirus pandemic may push humanity to realize the need for civilizational unity. One of the expected obvious consequences could be the appearance of the project of a world multilingual scientific alive encyclopedia. Such a project would be a worthy alternative to Wikipedia.

The mass migration of authors to alive publications would cause many difficulties for publishers, reviewers, editors, lawyers, librarians, and bibliometrists. It will require rethinking some basic concepts and reworking or redeveloping many of the software tools that support modern publications. However, the scale of these reforms is not so significant as to raise doubts about their speedy implementation. The main thing is that the results will benefit authors and readers, i.e., the principal actors in any publication.

Tools to support alive publishing are constantly being improved, and authors of scientific articles master the skills of handling online materials. Thus, maintaining an article in an alive publication mode is gradually becoming more uncomplicated and accessible.

Only absolute geniuses write perfect texts on the first attempt. All other authors will notice this or that imperfection of their publication after some time and will undoubtedly be happy to have a window of opportunity to improve, correct, or update it. A new paradigm for presenting the results of research is the future. Alive publications will replace many of the current forms of publications based on print traditions. In a few years, the scientist's mind will be transformed. Taking care to keep a publication up to date will become the norm; moreover, it will become a long-term, irresistible, and vital need, akin to a parent's care for a child's development.

## Contents

1. **Great Russian encyclopedia**
   ↑ **https://bigenc.ru**

2. **Russian-language Wikipedia**
   ↑ **https://ru.wikipedia.org**

3. **Living Reviews**
   ↑ **https://springer.com/gp/livingreviews**

4. **Living Evidence**
   ↑ **https://f1000research.com/collections/livingevidence/about-this-collection**

5. Fowler M.
   ↑ **EvolvingPublication**
   **https://martinfowler.com/bliki/EvolvingPublication.html**

6. Heller L., The R. and Bartling S.
   ↑ **Dynamic publication formats and collaborative authoring**
   *In: (2014) Bartling, S., Friesike, S. (eds) Opening Science. Springer, Cham.*
   **https://doi.org/10.1007/978-3-319-00026-8_13**

7. Simon J., Birukou A., Casati F., Casati R. and Marchese M.
   ↑ **Dynamic publication formats and collaborative authoring**

---

The tools mentioned above can be seen in action in the HTML version of this article posted in the institutional archive https://keldysh.ru/gorbunov/alive.